\documentclass[12pt]{article}
\input epsf.tex
\usepackage{cite}
\usepackage{latexsym}
\usepackage{graphicx}
\usepackage{amsmath,amstext}
\usepackage{comment}
\numberwithin{equation}{section}
\newcommand{\be}{\begin{equation}}
\newcommand{\ee}{\end{equation}}
\newcommand{\benn}{\begin{equation*}}
\newcommand{\eenn}{\end{equation*}}
\newcommand{\bea}{\begin{eqnarray}}
\newcommand{\eea}{\end{eqnarray}}
\newcommand{\bean}{\begin{eqnarray*}}
\newcommand{\eean}{\end{eqnarray*}}

\def\centeron#1#2{{\setbox0=\hbox{#1}\setbox1=\hbox{#2}\ifdim
\wd1>\wd0\kern.5\wd1\kern-.5\wd0\fi
\copy0\kern-.5\wd0\kern-.5\wd1\copy1\ifdim\wd0>\wd1
\kern.5\wd0\kern-.5\wd1\fi}}
\def\ltap{\;\centeron{\raise.35ex\hbox{$<$}}{\lower.65ex\hbox{$\sim$}}\;}
\def\gtap{\;\centeron{\raise.35ex\hbox{$>$}}{\lower.65ex\hbox{$\sim$}}\;}

\def\lsim{\mathrel{\ltap}}
\begin{document}
\begin{titlepage}
\begin{center}
\hfill SCIPP-2004/18\\

\vskip 0.2in

{\Large \bf The NMSSM, Anomaly Mediation and a Dirac Bino}

\vskip 0.3in

Linda Carpenter,$^1$ Patrick J. Fox,$^2$ and David E. Kaplan$^1$

\vskip 0.2in

\emph{$^1$ Johns Hopkins University\\
        Baltimore MD}

\vskip 0.1in

\emph{$^2$ Santa Cruz Institute for Particle Physics,\\
     Santa Cruz CA 95064}

\begin{abstract}
   We introduce a new model of supersymmetry breaking dominated by
   anomaly mediation.  It has a viable spectrum, successful electroweak
   symmetry breaking, solves the mu-problem and maintains the
   anomaly-mediated form for soft-masses down to low energies thus
   solving the flavor problem.  The model consists of the minimal
   supersymmetric standard model plus a singlet, anomaly-mediated soft
   masses and a dirac mass which marries the bino to the singlet.  We
   describe a large class of models in the UV which can produce such
   boundary conditions.  The dirac mass does not affect the so-called
   ``UV insensitivity" of the other soft parameters to running or
   supersymmetric thresholds and thus flavor physics at intermediate
   scales would not reintroduce the flavor problem.  The dirac bino is
   integrated out at a few TeV and produces {\it finite and positive}
   contributions to all hyper-charged scalars at one loop thus
   producing positive squared slepton masses.  The theory predicts some
   CP violation in the Higgs sector leading to a correlation between
   the spectra to be seen at the LHC and electric dipole moments within
   experimental reach in the near future.
\end{abstract}

\end{center}
\end{titlepage}

\section{Introduction}

Anomaly Mediated supersymmetry breaking (AMSB) \cite{Randall:1998uk}
(see also \cite{Giudice:1998xp}) is a very predictive
form of supersymmetry breaking with many desirable features, chief
amongst these is its insensitivity to intermediate scales (so-called
``UV insensitivity'').  Soft masses and scalar couplings are determined
by low energy couplings, independent of the details of the running
from the high scale.  Unfortunately this predictivity is also its
major drawback.  Sleptons, being charged under non-asymptotically free
gauge groups, have negative squared masses.  Attempts to fix this 
problem~\cite{Chacko:1999am,Katz:1999uw,Jack:2000cd,Carena:2000ad,Allanach:2000gu,Chacko:2000wq,Kaplan:2000jz,Arkani-Hamed:2000xj,Jones:2001iv,Abe:2001cg,Chacko:2001jt,Nelson:2001ji,Okada:2002mv,Nelson:2002sa,Jack:2003qg,Anoka:2003kn,Kitano:2004zd,Shafi:2004cf,Sundrum:2004un,Ibe:2004gh}
often result in reintroducing a dependence on the UV scale.  Here we
propose a modification of pure anomaly mediation that fixes the
slepton mass problem whilst retaining UV insensitivity.

Anomaly mediation also has a $\mu$-problem.  Including a $\mu$ term in
the superpotential explicitly breaks the conformal symmetry and
generates a B$\mu$ term that is a loop factor too large.  If instead
one works with the next-to-minimal supersymmetric standard model
(NMSSM) then conformal invariance is not explicitly broken at tree
level and it is possible to get correct EWSB.  So, we will work with
the NMSSM which includes a SM singlet, $S$.  The same
singlet will also be used to fix the tachyonic slepton problem in a UV
insensitive way.

The feature we add
to AMSB is Supersoft Supersymmetry breaking
(SSSB) \cite{Fox:2002bu}.  SSSB is a way of generating Dirac
gaugino masses which, when integrated out, produce finite positive 
squared scalar
masses a loop factor smaller than the gaugino mass.  SSSB is UV
insensitive since the scalar masses only run once the gauginos have
been integrated out.  In our model,
the bino marries a singlet and we show that this singlet can be the 
same singlet
that appears in the NMSSM.  We have a supersoft contribution for fields
charged under $U(1)_Y$ from a Dirac mass for (only) the bino.

The negative squared masses are generated at two loops by anomaly 
mediation
whereas the positive SSSB contributions are generated at one loop.
The scale of the AMSB contribution is set by the gravitino mass and the
SSSB contribution by the D-term vev of a hidden-sector $U(1)^\prime$.  
We will
demonstrate how these two scales arise in a model of dynamical SUSY
breaking and can be comparable.  The result is that the slepton
mass squareds are pushed positive and the spectrum becomes viable.

The rest of this paper is laid out as follows.  In Section
\ref{sec:review} we review AMSB and SSSB and set conventions for the
remainder of the paper.  We also discuss their mutual UV insensitivity. 
  We
then address the issue of EWSB, in Sections \ref{sec:nmssm} and
\ref{sec:spectrum}, and calculate the spectrum for the squarks and
sleptons.  We treat the relative sizes of the two contributions as a
free parameter and show that varying this parameter produces viable
spectra.  In Section \ref{sec:cpviolation} we discuss CP violation
and show that while CP is maximally broken in the Higgs sector, our
model lives at the boundary of current experimental probes.
In Section \ref{sec:uvmodels} we present possibilities for the UV
physics that reproduces our model.  For example, until this stage we
have treated the two supersymmetry-breaking scales as independent.  In
\ref{sec:fouronemodel} we discuss an explicit model of
dynamical SUSY breaking that relates these two scales and we show how
they typically have the right ratio.  In \ref{sec:dtermtransmission} we 
present
extra-dimensional realizations that naturally suppress all other 
sources of
supersymmetry breaking.  In \ref{sec:gutsymms} we address the issue of 
the
one remaining dangerous operator, kinetic mixing between hidden and 
visible
sector $U(1)$s, and discuss possibilities for natural suppression.
In Section \ref{sec:conclusion} we conclude.

\section{Anomaly and Supersoft Mediation}\label{sec:review}

In order for AMSB to dominate the hidden sector which breaks SUSY must
be sequestered from the MSSM.  This forbids contact interactions that
would otherwise dominate.  The sequestering was done originally via a
five-dimensional setup \cite{Randall:1998uk}, but recently was
realized entirely in four dimensions using a strongly coupled CFT
\cite{Luty:2001jh,Luty:2001zv}.  We will need the same sequestering
with the exception of couplings between a hidden sector gauge field
and visible sector fields.  This will be accomplished by putting the
hidden sector gauge fields in the bulk.  We will show explicitly in
Section \ref{sec:uvmodels} how to generate only the desired operators.

In such a set up the SUSY breaking is a result of the F-term of the 
conformal
compensator, $\Phi=1+\theta^2 m_{3/2}$, whose value is determined by 
tuning
the cosmological constant to zero.  After rescaling fields the conformal
compensator appears with the cutoff of the theory, as well as any 
explicit
mass scales in the Lagrangian.  When regulating loops this $\theta$ 
dependence
of the cutoff leads to superpartner masses.  For instance, consider the 
gauge
kinetic term,
\be
\int d^2\theta \frac{1}{g^2(\mu)} W_\alpha W^\alpha,
\ee
where the holomorphic gauge coupling, $g_h^2(\mu)$,  runs only at 
1-loop,
i.e.
\be
\frac{1}{g_h^2(\mu)}=\frac{1}{g_h^2(\Lambda)}+ b_0 \log
   \left(\frac{\mu}{\Lambda\Phi}\right).
\ee
The cutoff scale comes with powers of the conformal compensator, 
expanding the
logarithm leads to a mass for the gaugino,
\be\label{eq:gauginomass}
m_{\lambda_i}=\frac{\beta(g_i)}{g_i}m_{3/2}.
\ee
A similar calculation goes through for the wavefunction renormalization 
of the
chiral fields leading to soft masses and A-terms of,
\bea\label{eqn:ammasses}
m^2_i = -\frac{1}{4} \dot{\gamma}m_{3/2}^2 & & A_{ijk}= 
-\frac{1}{2}(\gamma_i
+\gamma_j +\gamma_k) m_{3/2}
\eea
where the dot corresponds to $d/dt$ and $t\equiv \log\mu$ identifies the
renormalization scale.  Here we see the UV insensitivity, these
results are true at all scales -- in order to calculate the masses in
the IR we need only know the value of couplings at that scale.
Unfortunately this predicts negative squared masses for any fields 
charged
under non-asymptotically free gauge groups.  In particular the
sleptons are tachyonic.  This is one of the shortcomings of AMSB.

SSSB requires an auxiliary $U(1)^\prime$ gauge symmetry which is
broken at a high scale where the auxiliary component of the vector
superfield gets a D-term vev.  Including the singlet in the MSSM makes
it is possible to write down supersymmetry-breaking operators,
\bea\label{eqn:supersoftops}
\lambda_{0}\int d^2\theta \sqrt{2} \frac{W^\prime_\alpha W_Y^\alpha 
S}{M} &\phantom{===} &
\lambda_{1} \int d^2
\theta \frac{W^\prime_\alpha W^{\alpha\prime} S^2}{M^2}.
\eea
The factor of $\sqrt{2}$ is to simplify normalisations below.
The first type of operator marries gauginos with adjoint fermions
giving Dirac masses.  The Dirac bino mass generated by
(\ref{eqn:supersoftops}) is,
\be\label{eq:binomass}
m_1=\lambda_{0}\frac{D^\prime}{M}\equiv \lambda_{0} m_D
\ee

As well as generating Dirac gaugino masses the first operator gives a
gaugino scale mass to the real part of the adjoint scalar leaving the
pseudoscalar massless\footnote{Here we take $\lambda_{0}$ real for
   simplicity.  Later we will allow for arbitrary phases when we
   discuss CP violation in Section \ref{sec:ewsb}.}.  It also generates
new scalar trilinear vertices involving the singlet scalar and charged
MSSM fields.  As a result, in addition to the usual one loop
corrections to scalar masses coming from gauge interactions there are
now new diagrams involving the singlet scalar.  These diagrams cut off
the loop integrals above the gaugino mass, resulting in scalar masses
that are finite and thus UV insensitive.  \emph{Scalar masses are only
   generated once the gauginos have been integrated out.}  A simple way
to see this is to note that there are no counterterms allowed by the
symmetries.  The lowest dimension operator which produces a scalar
mass for MSSM fields would be
\be
\int d^4 \theta \frac{W' W' W'^\dagger  W'^\dagger}{M^6} Q^\dagger Q ,
\label{scalar-higher-order}
\ee
which is proportional to $D'^4$ and is not a counterterm for divergent
contributions which would be proportional to $D'^2$.  The supersoft
scalar mass for a scalar of $U(1)_Y$ charge $Q$ is given by,
\be\label{eq:ssscalarmass}
m_{ss}^2=Q^2 \frac{\alpha_Y}{\pi} m_1^2 \log 
4(1+\lambda_{1}/\lambda_{0}^2).
\ee

The second operator in (\ref{eqn:supersoftops}) splits the real and
imaginary parts of the adjoint scalar, giving one a positive mass
squared and the other a negative mass squared.  Depending on the
relative sign and size of these two operators one component of the
scalar singlet may acquire a vev, this will be discussed further when
we discuss the breaking of electroweak symmetry.

We have described how both supersoft and AM are, by themselves, UV
insensitive.  There is still the question of whether in combination
they remain UV insensitive.  The gaugino now has both a Majorana mass
term generated by anomaly mediation and a Dirac mass term from
supersoft.  Above the Dirac mass the only contribution to the running
of scalar masses is from the gaugino's Majorana mass and this is
precisely the contribution necessary to keep the scalars running on
the anomaly mediated trajectory.  The Dirac and Majorana mass log
divergences at all loop orders renormalize independently due to an
R-symmetry.  Even (supersymmetric) thresholds to leading order in
$D'/M$ maintain the structure of the soft terms in the infrared, as
long as there are no particles which are charged under both $U(1)$s.
This is clear for scalars as the lowest order operator which can be
written is that in (\ref{scalar-higher-order}).

Thus, above the Dirac gaugino mass the two SUSY breaking mechanisms
are decoupled and the low energy physics is insensitive to all physics
above the Dirac mass scale\footnote{This was shown explicitly at the
one and two loop level in \cite{Jack:1999fa}.}.  At the Dirac mass
scale the bino is integrated out and generates a finite positive
mass squared for all scalars charged under hypercharge.  At this scale
($\mathcal{O}(10\,\mathrm{TeV})$) the running is pushed off the anomaly
mediated trajectory.  The combination of supersoft and anomaly
mediation is as insensitive as supersoft alone and the running above
the Dirac mass scale is purely that of AMSB.

As far as IR phenomenology is concerned there are two SUSY breaking 
mass scales
that determine superpartner masses at the Dirac mass scale, the AM scale
($m_{3/2}$) and the supersoft scale ($m_D$).  We will parametrize these 
as
$m_{3/2}$ and their ratio, $r=m_D/m_{3/2}$.  We will show later how 
these two
scales can be related through dynamics, making $r$ a derived quantity.  
When
we give examples of superpartner spectra we will ignore the small 
effects of
running from the bino mass scale.

\section{The NMSSM: model and spectra}\label{sec:nmssm}\label{sec:spectrum}

As discussed in Section~\ref{sec:review} if there are any explicit
mass scales in the superpotential they appear with powers of the
conformal compensator upon rescaling of fields.  Therefore in the MSSM
with a $\mu$ term, a $B\mu$ term will be generated of size $\mu
m_{3/2}$, much larger than the soft Higgs mass squared which is
$\mathcal{O}((m_{3/2}^2/16\pi)^2)$.  Instead we combine the NMSSM and
AMSB thus removing all renormalizable operators involving explicit
mass parameters.  
The relevant piece of the superpotential (including supersoft operators) is,
\be 
W=\lambda_S S H_u H_d + \frac{\kappa}{3}S^3+
\frac{\lambda_{1}}{M^2}W^\prime W^\prime S^2+
\frac{\lambda_0}{M}W^\prime W S + y_t Q H_u \bar{U}+\mathrm{h.c},  
\ee 
where we ignore the effects of all Yukawas other than the top Yukawa, a good
approximation at low $\tan\beta$.
The resulting potential contains both superpotential terms and soft
terms, $V=V_{susy}+V_{soft}$.  With,
\bea
\label{eq:vsusy}
V_{susy}&=&|\lambda_{S}H_u H_d+\kappa S^2|^2+|\lambda_S S H_d|^2 +|\lambda_S SH_u|^2
+\lambda_{1} m_D^2 (S^2+S^{*2})\nonumber \\
&& +\frac{1}{8} \left(g_Y(|H_u|^2-|H_d|^2)+4\lambda_{0} m_D
Re S\right)^2 +\frac{g_2^2}{8}
\left(|H_u|^2-|H_d|^2\right)^2\nonumber \\ 
&& + \frac{g_2^2}{2} \left|H^+_u
H_d^{0*}+H_u^0 H_d^{-*}\right|^2
\eea
and\footnote{Note $H_u H_d \equiv H_u^+ H_d^- -H_u^0 H_d^0$.}
\be
\label{eq:vsoft}
V_{soft}=m_{H_u}^2|H_u|^2 + m_{H_d}^2|H_d|^2 + m_S^2 |S|^2 +\lambda_S
A_\lambda (SH_u H_d + \mathrm{h.c})+
\frac{\kappa}{3}A_\kappa(S^3+\mathrm{h.c.}).
\ee
These soft parameters get contributions from both anomaly mediation and
supersoft, in the case of Higgs soft masses, and purely anomaly mediation, for
A-terms and singlet masses.  The AMSB contributions are given by
(\ref{eqn:ammasses}) specialised to the case for our superpotential,
see Appendix A.

In Section~\ref{sec:spectrum} we will evaluate the
soft masses in the IR and give examples of UV parameters that result in a
viable spectrum.  We can limit the parameter space we need to analyse by
requiring that we get correct electroweak symmetry breaking.  

We now calculate the superpartner spectrum in the model.  The effect
of the supersoft operators (\ref{eqn:supersoftops}) is to give the
bino a large Dirac mass and to add a positive, finite contribution to
the squared mass of all fields, proportional to their hypercharge
squared.  There are also positive supersoft contributions to the
Higgses and singlet scalar masses through their superpotential
couplings.

In pure AMSB the superpartner masses can be calculated using
(\ref{eq:gauginomass}) and (\ref{eqn:ammasses}) and the formulae in
Appendix A.  The one-loop supersoft corrections are given by
(\ref{eq:binomass}) and (\ref{eq:ssscalarmass}), and for simplicity we
normalise $\lambda_{0}$ to $1$.  The relative size of the supersoft
and the AMSB contributions is set by the ratio of $r\equiv
m_D/m_{3/2}$.  In Section \ref{sec:uvmodels} we will demonstrate how
it is possible to dynamically generate $r\lsim 1/2$ but for now we
leave $r$ as a free variable.

To see how the addition of supersoft fixes the tachyonic slepton
problem we list the masses of the gauginos, the sleptons,
the down type squarks from all generations and the up type squarks
from the first two generations.  The soft masses of these fields
depend only upon gauge couplings whereas the Higgs scalar and
third generation squark masses depend on the particular choice of 
$y_t$, $\kappa$, and $\lambda_S$.
 
The winos, gluinos and bino/singletino system have masses
\bea
M_1 &=& \begin{pmatrix} 1.40\,M & 158\, r M \\ 158\, r M&  
\kappa |\langle S\rangle| 
  \end{pmatrix}  \\
M_2 &=& 0.427\, M \\
M_3 &=& -4.32\, M,
\label{eq:gauginomatrix}
\eea
where $M=m_{3/2}/(16\pi^2)$.  The scalar masses that are independent
of the top Yukawa are
\bea
m_{\tilde{L}_{1,2,3}}^2 &=&   (-0.363  +  28.1\,  r^2) M^2\\
m_{\tilde{e}_{1,2,3}}^2 &=&   (-0.358  + 112\,    r^2) M^2\\
m_{\tilde{Q}_{1,2}}^2 &=&   (16.3    +   3.12\, r^2) M^2\\
m_{\tilde{u}_{1,2}}^2 &=&   (16.4    +  49.9\,  r^2) M^2 \\
m_{\tilde{d}_{1,2,3}}^2 &=&   (16.5    +  12.5\,  r^2) M^2.
\eea
It is clear that a large enough $r$ will fix the tachyon problem.
From (\ref{eq:ssscalarmass}), we see that the exact size of the
supersoft correction depends not only upon $r$ but also
$\lambda_{1}$.  However, as we will see later at viable points
$\lambda_{1}\ll 1$ so the $\log$ in (\ref{eq:ssscalarmass}) is well
approximated by $\log\,4$,we have used this approximation when
calculating the masses above.

In (\ref{eq:gauginomatrix}) we have included the Majorana mass for the
singletino that arises, when $S$ gets a vev, from the cubic term in
the superpotential.  We ignored this effect when calculating the
supersoft contribution (\ref{eq:ssscalarmass}), the correction is
small--of order $\kappa^2 \langle S\rangle^2/m_D^2$.  In addition,
there is a small Majorana mass for the gaugino coming from AMSB.  Both
Majorana masses are much smaller than the Dirac mass and we ignore the
small splitting in the bino-singletino system, quoting just the Dirac
mass when we give spectra.

All that remains is to find appropriate values for the parameters
$\lambda_{S}$, $\lambda_{1}$, $\kappa$, $r$ and the one mass scale
$M$ that lead to superpartner masses consistent with present
experimental constraints and viable EWSB.

\subsection{EWSB}
\label{sec:ewsb}

The Higgs-singlet potential that must break electroweak symmetry is
given by (\ref{eq:vsusy}) and (\ref{eq:vsoft}).  The singlet vev is
controlled by the $\lambda_1$ coupling (having normalised $\lambda_0$
to 1).  We now describe what happens as $\lambda_1$ is varied.

We first consider $\lambda_1<0$.  For $-3/4<\lambda_1<0$ the vev of
$S$ is tiny, consequently so is the $\mu$-term, and the supersoft
corrections to the slepton masses are positive,
(\ref{eqn:supersoftops}).  At $\lambda=-3/4$ the supersoft
contributions turn off.  For $-1<\lambda_1<-3/4$ the supersoft
contributions change sign since the scalar $S$ is now heavier than the
bino.  As we approach $\lambda_1=-1$ the mass of the real part of $S$
goes to zero and the slepton mass contributions go to $-\infty$, an IR
divergence.  Lowering $\lambda_1$ further increases the supersoft mass
from $-\infty$, to 0 at $\lambda_1=-5/4$, at the same time the mass
for the real part of $S$ decreases from 0 to $-m_D^2$ leading to a
large real $S$ vev.  Below $\lambda=-5/4$ the supersoft contributions
are again positive but the vev of the real part of $S$ is greater than
$m_D$.  However, the appearance of $S$ in the hypercharge D-term
forces the Higgses to acquire unacceptably large vevs.  We see that
the whole region of $\lambda_1<0$ is ruled out, leaving only positive
$\lambda_1$.

The region $\lambda_1>0$ gives reasonable supersoft slepton masses and
a negative squared mass for the imaginary part of $S$, of magnitude
$\lambda_1 m_D^2$.  The vev of $S$ determines the $\mu$ and $B\mu$
terms and so we take $\lambda_1$ small.  However, it can not be taken
arbitrarily small due to the fact that we now have a CP violating vev
and too small a $\lambda_1$ leads to a large contribution to EDMs,
discussed below.  From now on we consider $\lambda_1$ small and positive.

While the Higgs vev of $175$ GeV and a reasonable $\mu$ term are relatively
easy to achieve by the correct choice of the overall scale of the
potential, one issue that persists over a very large range of
parameters is that $\tan\,\beta$ is low ($\sim 1$) and Im$\langle
H_d\rangle\approx \mathrm{Re}\, \langle H_d\rangle$ since the
potential is relatively symmetric with respect to both the real and
imaginary parts of the field.  The value $\tan\,\beta\sim 1$ is not a
problem for the physical Higgs mass, but requires a too-large top
Yukawa coupling if one wants to avoid Landau poles below the GUT
scale.

However, there is one more operator that can be added to the Higgs
potential,
\be 
\frac{\lambda_c}{M^2} W^\prime W^\prime H_u H_d 
\ee 
When the $W^{'}$ gets a vev, this is a $B\mu$ term for the Higgs which
is $\lambda_c m_D^2$. This term also splits the real and imaginary
parts of the vev of $H_d$, if we allow the coupling constant $\lambda_c$ to
be complex.  

There is a portion of parameter space in which $\mu$ is of moderate
size ($\sim 100$ GeV) and the large negative contribution to the
(up-type) Higgs soft mass is partly canceled by a large positive
contribution from the finite bino loop ({\it i.e.}, the supersoft
contribution), and the new B$\mu$ term.  In this range (see Table
\ref{tab:points} below) it is possible to get a reasonably-sized tan
$\beta$ and a viable spectrum.  Even the (standard-model-like) Higgs
mass is typically well above the LEP bound.
\begin{table}
\begin{center}
\begin{tabular}{c|c|c|c|c}
& & Point 1 & Point 2 & Point 3\\
\hline
inputs:  &
$\lambda_{1}$& $5.0\times 10^{-3}$ & $7.7\times 10^{-3}$ & $10\times 10^{-3}$\\ 
&$-\lambda_c$& $3.0\times 10^{-4}$ & $1.3\times 10^{-4}$ &  $1.0\times 10^{-4}$\\
&$\lambda_S$& $2.3\times 10^{-2}$ & $1.6\times 10^{-2}$ & $1.7\times 10^{-2}$\\
&$\kappa$      & 0.28  & 0.34 & 0.36\\
&$y_t$         & 1.135 & 1.14 & 1.15\\
&$r$           & 0.487 & 0.46 & 0.45\\
&M             & 443   & 755  & 506\\ 
\hline
sleptons:
&$m_{\tilde{e}_L}$  & 1110 & 1780 & 1170\\
&$m_{\tilde{e}_R}$  & 2270 & 3650 & 2400\\
&$m_{\tilde{\nu}_L}$& 1110 & 1780 & 1170\\
\hline
squarks:
&$m_{\tilde{u}_L}$ & 1830 & 3110 & 2080\\
&$m_{\tilde{u}_R}$ & 2350 & 3920 & 2610\\
&$m_{\tilde{d}_L}$ & 1830 & 3110 & 2080\\
&$m_{\tilde{d}_R}$ & 1960 & 3310 & 2210\\
\hline
stops:
&$m_{\tilde{t}_1}$ & 1740  & 2950 & 2000\\
&$m_{\tilde{t}_2}$ & 2210  & 3670 & 2450\\
\hline
gauginos:
&$m_{\tilde{B}}$ & 34100 & 54900 & 36000\\
&$m_{\tilde{W}}$ & 189   & 322   & 216\\
&$m_{\tilde{g}}$ & 1910  & 3260  & 2190\\
\hline
Higgs sector:
&$|\mu|$         & 159     & 184     & 143\\
&$tan\,\beta$    & 2.42    & 3.18    & 2.50\\
&Arg\,$\mu$      & $\pi/2$ & $\pi/2$ &  $\pi/2$\\
&Arg$\,\langle H_u H_d\rangle$& 0.116$\pi$ & 0.124$\pi$ & 0.124$\pi$ \\
&$m_{h^0}\,$&
106 & 114 & 108 \\
\hline
neutron edm: & $|d_n|$(e cm) & $3.24\times 10^{-26}$ & $6.0\times 10^{-27}$
& $1.2\times 10^{-26}$\\
electron edm: & $|d_e|$(e cm) & $1.23\times 10^{-27}$ & $4.8\times 10^{-28}$
& $1.0\times 10^{-27}$
\end{tabular}
\caption{Typical solutions, all masses in GeV.  Here $m_{\tilde{B}}$
  denotes to the Dirac mass between the bino and the singletino.}
\label{tab:points}
\end{center}
\end{table}

There are several general features worthy of note in
Table~\ref{tab:points}.  The mass of the lightest Higgs quoted in the
table includes the dominant loop corrections and has been
calculated~\footnote{We thank Carlos Wagner for assistance with this.}
for a top quark pole mass of 178 GeV but it is undoubtedly subject to
further corrections.  There is spontaneous CP violation so the Higgs
states can not be split into a CP-even and a CP-odd sector -- they
mix.  The singlet state is heavy and the low energy Higgs
phenomenology has similar aspects to that of the MSSM with explicit CP
violation \cite{Carena:2002bb}.  This CP violation predicts electric
dipole moments not far beyond present experimental bounds.  The LSP is
a neutralino, a linear combination of a neutral Higgsino and Wino.

For our model, $y_t$ is typically the most
sensitive parameter.  Normally we would define the sensitivity
as $\partial \log v/\partial \log y_t$.  However, our points lie on
the edge of parameter space; while decreasing $y_t$ produces only a
moderate change in the Higgs vev, increasing $y_t$ by a minute amount
causes a very large change.  
Instead we define a parameter comparing the region of parameter space
over which $y_t$ assumes it's natural value, where the Higgs vev wants to
live around the scale $m_{3/2}$, to the region on the edge of parameter
space where the Higgs vev is around\footnote{More precisely we find
  the region of parameter space where the Higgs vev varies by at most
  100\% from the Standard Model value.} 175GeV.  For our points, this
naturalness parameter is of order $10^{-3}$.
  
We have attempted to use the singlet already present in the NMSSM and
an integral part of EWSB to also solve the tachyon problem of AMSB.
This is the minimal model and it turns out to be tightly constrained,
as can be seen from the fraction of a percent the naturalness
parameter assumes.  We will discuss in the conclusion possible ways to
alleviate this tuning.


\section{CP Violation}
\label{sec:cpviolation}

In order to calculate the amount of CP violation predicted by the
model we must first identify all the CP violating phases.  In
principle many parameters in the model could contain a phase but
because the soft parameters are generated through anomaly mediation
many of these phases are related.  In particular, the Majorana gaugino
mass ($m_i$) and all the A-terms ($A_h$, $A_\lambda$ and $A_\kappa$)
are proportional to $m_{3/2}$ with no new phase entering through
the running.  In addition to the phases present in parameters there is
also spontaneous CP violation coming from the vevs for $S$, $H_u$
and $H_d$.  

The simplest way to identify the physical CP violating phases is to
notice that the superpotential and soft parameters break three $U(1)$
symmetries.  By allowing both fields and parameters to transform we
can restore the symmetries.  The charge assignments are
given in Table~\ref{tab:charges}.  The physical CP violating phases will be
invariant under the spurious symmetries.

The superpotential generically has five couplings with phases \footnote{Note
  that we are only interested in phases beyond those already present
  in the SM, so we ignore the QCD vacuum angle as well as the Yukawas
  and their associated CKM phase.}  
\be
\lambda_{0},\ 
\lambda_1,\  
\lambda_c,\ 
\lambda_S,\ 
\kappa.
\ee
In addition, there are four phases that come from the soft SUSY breaking
terms, 
\be
m_i,\ 
A_y,\ 
A_\lambda,\ 
A_\kappa,
\ee
all of which have the same phase, as was explained above.  Phases can
also come from the vevs of $S$, $H_u$, and $H_d$ but one combination of
these phases can be removed by an $SU(2)$ gauge transformation.  We
have a total of eight phases, however due to the symmetries only five
are physical CP violating phases.  
\vspace {0.1in}
\begin{table}
\begin{center}
\begin{tabular}{|c|ccc|}
\hline
  & $U(1)_{PQ}$ & $U(1)_{R^\prime}$ & $U(1)_S$ \\
\hline
$\lambda_0$& 0& $0$& -1 \\ 
$\lambda_{1}$& 0& $0$& -2 \\ 
$\lambda_S$& -2& 2& -1 \\ 
$\lambda_c$& -2& 0 & 0 \\ 
$\kappa$& 0& 2& -3 \\
$m_i,\, A_{y,\lambda,\kappa}$& 0& -2& 0 \\ 
$H_{u,d}$& 1&0 & 0 \\ 
$S$& 0& 0& 1 \\ 
\hline
\end{tabular}
\caption{U(1) charges of parameters and fields.}
\label{tab:charges}
\end{center}
\end{table}

There are three independent, CP violating combinations of parameters allowed by
the charge assignments in Table~\ref{tab:charges},
\be
\label{eq:cpviolparams}
\lambda_0^2\lambda_{1}^*,
\  \lambda_0^3A_i^*\kappa^*,\  
\lambda_0 \lambda_S^*A_i^*\lambda_c,
\ee
where $A_i$ can be any of $A_\kappa$, $A_\lambda$, $A_y$ and $m_i$.
Including vevs of fields, we can write down two more combinations
which are both invariant under the $U(1)$'s above and also gauge
invariant,
\be
\label{eq:cpviolvevs}
\lambda_0\langle S\rangle\  \mathrm{and}\  \lambda_S A_i 
\langle S\rangle \langle H_u H_d\rangle .
\ee
Thus any CP violating physical amplitude must be a function only of the
combinations of parameters and vevs given in (\ref{eq:cpviolparams})
and (\ref{eq:cpviolvevs}).  At the one-loop level we will see this explicitly.

The strongest constraints on new CP violating phases come from
attempts to measure electric dipole moments (EDMs) of the electron and
the neutron.  
The effective dipole operator coupling fermions to the photon is,
\be
-\frac{1}{2}\mathcal{D}_\psi
\overline{\psi}\,\sigma^{\mu\nu}\psi\,\gamma_5\,F_{\mu\nu} + h.c.
\label{eq:edmop}
\ee
The coefficient of the operator, $\mathcal{D}_\psi$, is related to the
electric dipole moment, $d_\psi$ as \cite{Graesser:2001ec,Abel:2001vy}, 
\be
\label{eq:dipoleop}
d_\psi=|\mathcal{D}_\psi|\sin\phi\ \ \ \mathrm{with}\ \
\phi=\mathrm{Arg}(m_\psi^* 
\mathcal{D}_\psi). 
\ee
For the electron the present bound on its EDM is derived from the
measurement of the EDM of $^{205}$Tl and is \cite{Regan:2002ta} $|d_e|\leq
1.6\times 10^{-27} e$ cm, the corresponding limit for the neutron
\cite{Harris:1999jx} is $|d_n|\leq 6.3\times 10^{-26} e$ cm both with
90\% confidence.  

First the case of electron EDMs. On the face of it this operator
appears to be dimension 5 but it is chirality violating and so is
proportional to the mass of the fermion, making it effectively
dimension 6.  At leading order in the electron Yukawa one-loop
diagrams involving superpartners contribute to the EDM, for a review
of the calculation in the MSSM see, for example,
\cite{Graesser:2001ec} and \cite{Abel:2001vy} and references therein.
The calculation in our case is simplified because diagrams involving the bino 
are now suppressed since the Dirac mass of the bino is order many
TeV.  In the NMSSM there is an additional contribution but it again involves
exchange of a bino and can be ignored.
Using the physical points from Table~\ref{tab:points} as a
guide to the form of the spectrum we find,
\be
\mathcal{D}_e\sim \frac{g_2^2}{16\pi^2}m_e\left(\frac{2}{|m_{\tilde{l}}|^2}
+\frac{v_u^*}{v_d}\frac{
  m_{\tilde{H}}^* m_{\tilde{W}}^*}{|m_{\tilde{l}}|^4}\right),
\ee
and
\be
\phi=\mathrm{Arg}
\left(2 
+\frac{v_u^* m_{\tilde{H}}^* m_{\tilde{W}}^* 
}{v_d
  |m_{\tilde{l}}|^2}\right).
\ee
As promised, this phase is only a function of those
combinations of parameters and vevs (\ref{eq:cpviolparams}),
(\ref{eq:cpviolvevs}) allowed by the spurious symmetries.  We list the
resulting electron EDMs for each point at the bottom of
Table~\ref{tab:points}; they are close to but below the experimental
bound.  There are similar diagrams that contribute to quark EDMs which
in turn lead to a neutron EDM.  As for the leptons we are close to
but below the neutron EDM bound.

In addition, the neutron EDM receives a contribution from the
dimension 6 operator first discussed by Weinberg \cite{Weinberg:1989dx},
\be
-\frac{1}{6}C\, f_{abc}\,G^a_{\alpha
  \mu}\,G^{b\mu}_{\beta}\,G^c_{\gamma\delta}\,
\epsilon^{\alpha\beta\gamma\delta},
\label{eq:GGG}
\ee
where $f_{abc}$ are the structure constants.  This operator is not
proportional to a light quark mass and so is the dominant contribution
to the neutron EDM, it is not present for the electron as it requires
the fermion to be charged under a non-abelian gauge group.  The
largest contribution to the coefficient $C$ comes from a two loop
diagram containing 3 external gluons with the quarks in the loop
exchanging a Higgs boson.  For this contribution to be non-zero the
Higgs sector must have mixing between at least three scalar fields
e.g. the NMSSM.

Using naive dimensional analysis the effects of (\ref{eq:GGG}) on the
neutron EDM operator of (\ref{eq:edmop}) can be estimated.  The
dominant contribution requires two mass insertions on the Higgs
propagator. If the scalar states inserted are too light, the
contribution to the neutron EDM could be huge.  Fortunately we see from
electroweak symmetry breaking that the singlet mass is multi TeV and
the coupling $\lambda_S$ is small, of order $10^{-2}$, and the
contribution to the neutron EDM (see Table~\ref{tab:points}) is below,
but relatively close to the experimental bound.


\section{UV Models}\label{sec:uvmodels}

We now discuss several natural UV realisations of our model that result in the
interesting IR physics discussed above.

\subsection{Symmetries, Forbidden Operators and GUTs}
\label{sec:gutsymms}

Before we discuss the mediation of supersymmetry breaking we comment
on the remaining relevant operators that can affect weak scale
physics.

The most dangerous operator\footnote{We thank Hitoshi Murayama for
  reminding us of this fact.} which could be introduced is kinetic
mixing between the hidden sector $U(1)$ and hypercharge, {\it i.e.},
\be
\int d^2 \theta W_Y^\alpha W'_\alpha .
\ee
If this term is present the D-term for the hidden sector $U(1)$ would
be a tadpole for the hypercharge D-term, destabilising the gauge
hierarchy.  While there is no symmetry that could forbid this operator
as the theory stands, it is technically natural not to include this
term at the cutoff as it won't be generated through loops since there
is no matter in the theory that is charged under both groups.

Ultimately one would like to embed the whole model in a GUT, such as
$SU(5)$ \cite{Chacko:2004mi}, and doing so may allow us to suppress or
forbid entirely the kinetic mixing due to symmetries.  The GUT would
be broken down to $SU(3)\times SU(2)\times U(1)$ by an adjoint chiral
field, $\Sigma$, acquiring a GUT scale vev.  In addition to using this
field to break the GUT group we use it to pick out the $U(1)_Y$
direction so that the only supersoft operator is for $U(1)_Y$.  At the
GUT scale the supersoft operator of (\ref{eqn:supersoftops}) becomes, 
\be 
\int d^2\theta \sqrt{2} \frac{W^\prime_\alpha W^\alpha \Sigma S}{M^2}.  
\ee
The low energy supersoft operator would then be suppressed by a factor
of $\langle\Sigma\rangle/M$ which could be $\sim 10^{-2}$ or so, but
of course depends on the physics at the GUT/string/Planck scales.

In the context of a GUT, one could potentially use a symmetry to
forbid the kinetic mixing.  One possibility is to impose a $U(1)$
R-symmetry which is spontaneously broken by the GUT.  The singlet and
all the MSSM fields would have charge $2/3$ under the R-symmetry while
$\Sigma$'s charge would be $-2/3$.  This forbids the kinetic mixing
term even below the GUT scale, whilst allowing all the previous
operators discussed once appropriate powers of $\Sigma/M$ are
included.  However, it is an open question how to arrange for a field
carrying non-zero $U(1)_R$ charge to have a GUT breaking
(super)potential.  More importantly, there will be a modulus produced,
the superpartner of the R-axion, that could cause the GUT scale to act
effectively as a non-supersymmetric threshold and take us off the AMSB
trajectory \cite{Pomarol:1999ie,Katz:1999uw,Chacko:2000wq}.  Instead
of a continuous symmetry one could charge fields under a discrete
R-symmetry by adding terms in the GUT theory which explicitly break
the $U(1)_R$ but preserve the discrete subgroup.  As an example,
$\mathcal{Z}_{12}$ with charges -6 for $d^2 \theta$, +2 for $H_u$,
$H_d$ and $S$, +3 for $W_\alpha$ and $W_\alpha^\prime$, and -2 for
$\Sigma$.  The kinetic mixing term would thus be suppressed by a
factor of $(\langle\Sigma\rangle/M)^6$.  

\subsection{Supersymmetry Breaking: The $4-1$ Model}\label{sec:fouronemodel}

What we want is a situation where the $D$-term breaking in the $U(1)$
is comparable to the overall scale of supersymmetry breaking in the
hidden sector.  This can happen in any supersymmetry-breaking sector
in which the $U(1)$ is {\it required} for supersymmetry breaking, {\it
  i.e.}, if the $U(1)$ gauge coupling is turned off, a
supersymmetry-preserving minimum is restored.  While there are other
examples in the literature of this kind, we will describe a
particularly simple one \cite{Dine:1996ag}.

The model has an $SU(4)\times U(1)$ gauge group.  The matter content
consists of the following $SU(4)$ representations: an antisymmetric
tensor $A_2$, a fundamental $F_{-3}$, an anti-fundamental
$\bar{F}_{-1}$ and a singlet $S_4$.  The subscripts are the charges
under the $U(1)$.  The only allowed superpotential term is
\be
W=\lambda S_4 F_{-3}\bar{F}_{-1}
\ee
However $SU(4)$ will confine and the gauginos will condense leading to a
non-perturbatively generated superpotential.  The complete superpotential
becomes,
\be
W=\lambda S_4 F_{-3}\bar{F}_{-1} + \frac{\Lambda_4^5}{\left(\bar{F}_i
    F^jA^{ik}A^{lm}\epsilon_{jklm}\right)^{1/2}}.
\ee
We consider a regime where $g_4\gg\lambda\gg g_1$.  The minimum of the
potential will occur along the $SU(4)$ D-flat direction.  We can make a gauge
rotation so that the vevs along the $SU(4)$ D-flat directions have the form,
\be
A_2=\left(\begin{matrix} a\sigma_2 & \ \\ \ & a\sigma_2\end{matrix}\right),
F=\bar{F}=\left(\begin{matrix} b\\0\\0\\0\end{matrix}\right), S=c.
\ee
For convenience we rescale the fields, $\phi\rightarrow
\frac{\Lambda}{\lambda^{1/5}} \phi$.  In this paramatrization the
  $U(1)$ D-term is,  
\be
D_1=g_1\frac{\Lambda^2}{\lambda^{2/5}}(2|a|^2-4|b|^2+4|c|^2)
\ee 
whilst the contribution to the potential is $V_D=\frac{D^2}{2}$
and the F-term contribution to the potential is,
\be
V_F=\lambda^{6/5}\Lambda^4 \left(|b|^4+\left|2 b c -\frac{1}{ab^2}\right|^2 +
\left|\frac{1}{a^2 b}\right|^2\right)
\ee
The minimum of the potential has non-zero D-term.  If the $U(1)$ gauge
coupling is turned off it is possible to satisfy the F-term constraint,
$b\rightarrow 0$ while $a$ and $c$ scale like inverse powers of $b$
and SUSY is unbroken.  If we turn on the gauge coupling the minimum of the
potential moves in from infinity.  In order to get a non-zero D-term it is
crucial that the gauge group be involved in the SUSY breaking.  Numerically
we find that for $\lambda=10 g_1$ $V_D\approx 0.5 V_F$.  Thus, we have a
non-perturbative mechanism whereby the D-term and F-term contributions to the
vacuum energy are comparable.

The model above is not unique.  
There are other supersymmetry breaking models where a gauged $U(1)$
plays an important role in the dynamics, see for example \cite{Dine:1996ag}.

\subsection{D-term transmission}\label{sec:dtermtransmission}

Now we describe how to transmit supersymmetry breaking in the form of
anomaly mediation plus a dirac bino mass while sequestering all other
contributions.  This can be done via a fifth dimension
\cite{Randall:1998uk} or via conformal sequestering
\cite{Luty:2001jh,Luty:2001zv,Nelson:2000sn} with the hidden sector
$U(1)$ in the ``bulk''.

\subsubsection{Flat Extra Dimension}
\label{sec:sigmaflat}

We place the $SU(4)$ gauge fields and all of the matter of the 
4--1 model on the hidden brane while the
MSSM matter and gauge fields are restricted to lie on a separate brane at the
other orbifold fixed point.  The $U(1)$ gauge field of the 4--1
model propagates in the bulk.  The dynamics discussed in the previous
section generate a non-zero D-term on the hidden brane, $D_b$.  

In the bulk the vector multiplet is part of an $N=2$ gauge multiplet
that decomposes in $N=1$ language as a vector multiplet and a chiral
multiplet, also in the adjoint of the gauge group
\cite{Arkani-Hamed:2001tb}.  In Wess Zumino gauge,
\bean
V^\prime&=&-\theta \sigma^\mu \bar{\theta}A^\prime_\mu + i
\bar{\theta}^2\theta\lambda_1 - i \theta^2\bar{\theta}\lambda_1
+\frac{1}{2} \bar{\theta}^2\theta^2 D^\prime\\
\Phi&=& \frac{1}{\sqrt{2}}(\Sigma+iA_5)+\sqrt{2}\theta\lambda_2
+\theta^2 F.
\eean
We take the vector superfield to be even under the orbifold boundary conditions
and the chiral superfield to be odd.  The relevant part of the action is,
\bea
&&\int d^4 x dy \left[\int d^2\theta \frac{1}{4}W^{\prime\alpha}
  W^\prime_\alpha + h.c. + \int d^4\theta (\partial_5
  V^\prime-\frac{1}{\sqrt{2}}(\Phi+\Phi^\dagger))^2\right.\nonumber \\
&&\left. + \delta(y-\pi R) \int d^4\theta X^\dagger e^{g_5 V^\prime} X +
\delta(y) \int d^2\theta \frac{W^{\prime\alpha}W_{Y\alpha} S}{M^{3/2}}\right].
\eea
This leads to equations of motion
\be\label{eqn:Deom}
D^\prime +\partial_5 \Sigma+\delta (y-\pi R)\sqrt{2\pi R} D_b +
\delta(y) \frac{D_Y S}{M^{3/2}} =0 \ \ \mathrm{and}\ \  
\frac{1}{2}\eta^{\mu\nu}\partial_\mu\partial_\nu\Sigma -\partial_5 D^\prime=0.
\ee

In the bulk the zero mode of the scalar, $\Sigma$, has constant slope.
Without the boundary source terms the zero mode of $\Sigma$ would be
projected out and $\Sigma$ would be zero across the whole space.
Here, however, there are non-zero source terms at the branes causing
$\Sigma$ to jump at the branes.  Taking $D_Y=0$ there is only a jump
at $y=\pi R$, this causes $\Sigma$ to have a gradient across the whole
space.  Although its value at $y=0$ is unaltered from the case with no
sources it now has gradient there, leading to a non-zero D-term at
$y=0$, Figure~\ref{fig:sigmaflat}.  Using the periodicity of $\Sigma$
we find,
\be
D^\prime=-\frac{1}{\sqrt{2\pi R}}D_b 
\ee
and
\be
\Sigma=\frac{D_b}{\sqrt{2\pi R}}(y-2\pi R\  \theta(y-\pi R)),
\ee
where $\theta(y)$ is the Heaviside function.

A concern one should have raised by now is the fact that there are
light fields in the bulk (the $U(1)$ gauge multiplet) which could now
in principle communicate supersymmetry breaking directly to the
visible sector by generating scalar masses through higher dimensional
operators involving bulk fields.  The gauge invariant combination of
bulk fields which could couple to the visible sector is $(\partial_5
V^\prime-\frac{1}{\sqrt{2}}(\Phi+\Phi^\dagger))$.  This combination
has odd boundary conditions, but could couple through a partial
derivative, as
\be
\delta(y) \int d^4\theta \frac{\partial_5 (\Phi+\Phi^\dagger -
  \sqrt{2} \partial_5 V^\prime)}{M^{5/2}} Q^\dagger Q.
\label{eq:dangerous}
\ee
A scalar mass for $Q$ would come from the $D$-component of the
coefficient and thus proportional to $\partial_5^2 D^\prime$.
However, $D^\prime$ is constant in the bulk and through $y =0$, so
this potentially dangerous contribution vanishes {\it dynamically}.
In fact all such contributions cancel because gauge invariance
requires the gauge field to appear as $\partial_5 V^\prime$.

So the D-term on the hidden brane is transmitted through the bulk to our
brane.  Any F-term generated on the hidden brane is still sequestered from the
MSSM, forbidding SUSY breaking contact interactions.

\subsubsection{Warped Extra Dimension}
\label{sec:sigmawarped}

\begin{figure}[t]
\centerline{\epsfxsize=2in \epsfbox{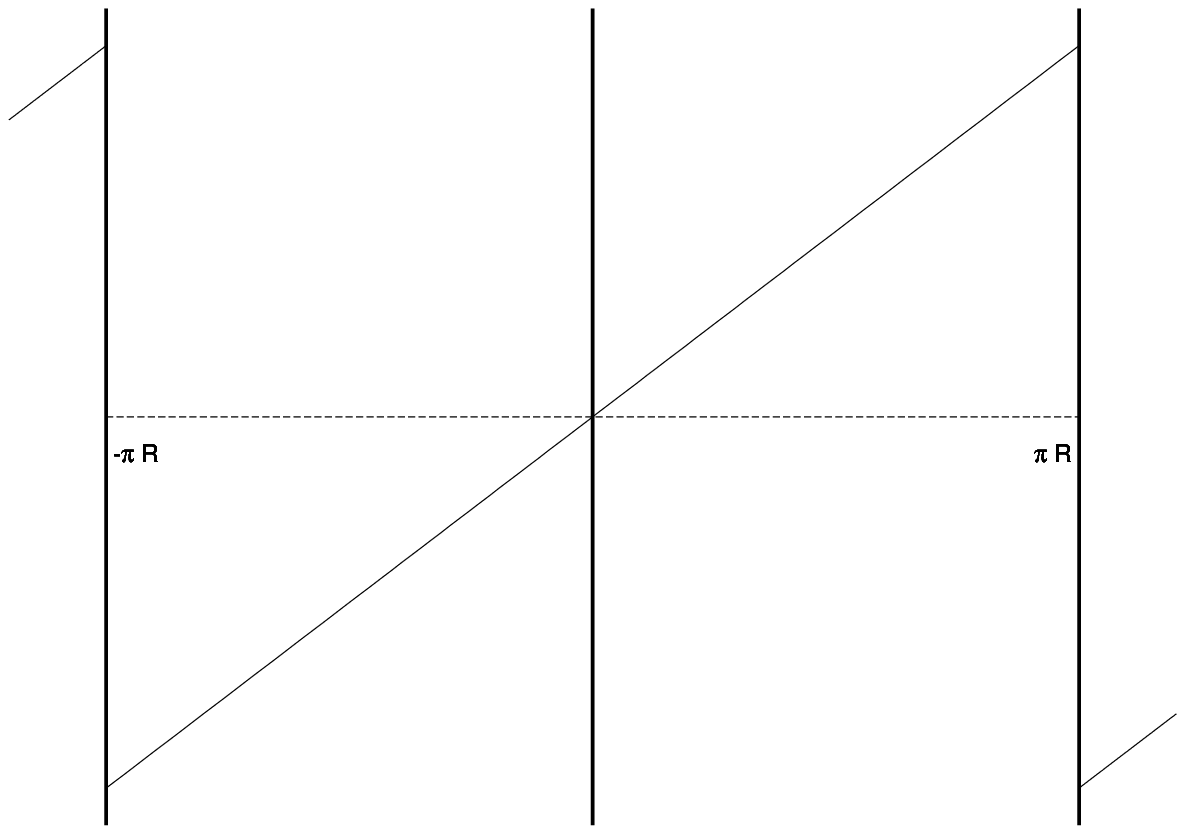}
\epsfxsize=2in \epsfbox{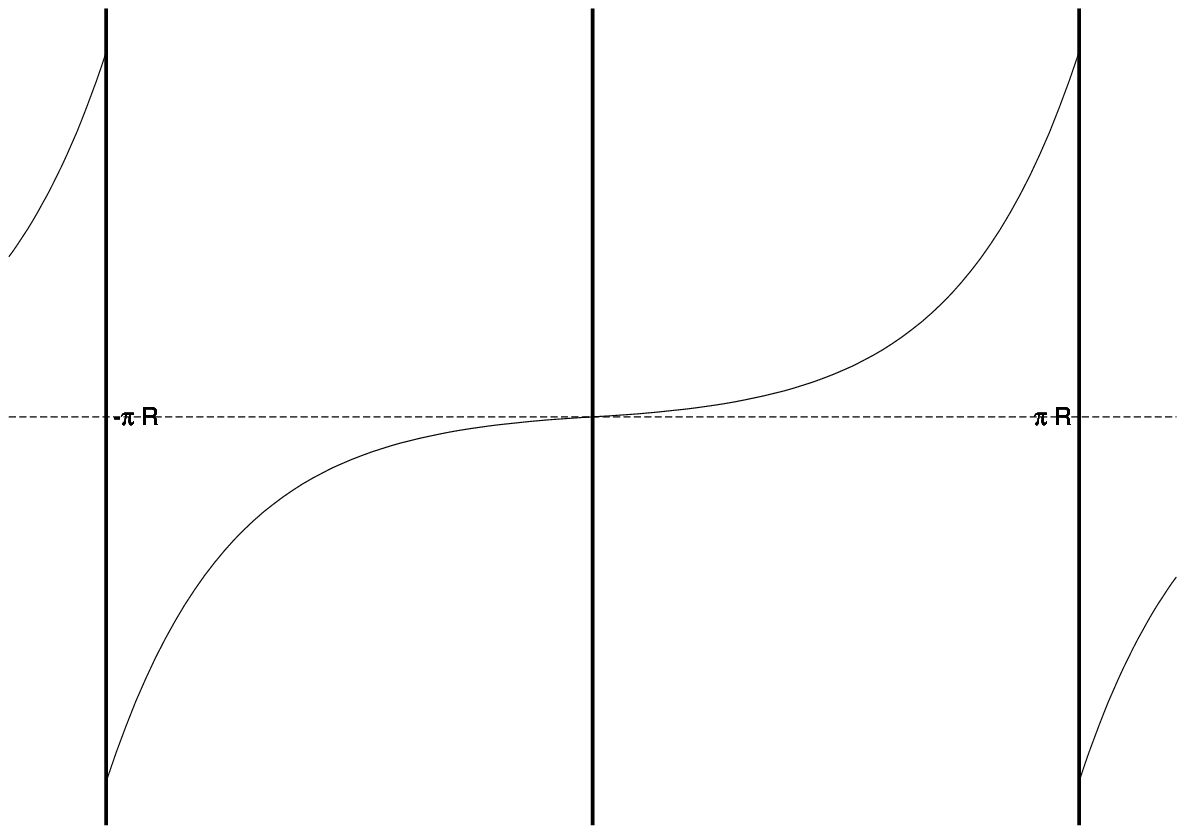}}
\caption{The profile of $\Sigma$ is shown for a flat extra
  dimension on the left and for the warped case on the right.}
\label{fig:sigmawarped}\label{fig:sigmaflat}
\end{figure}

We now wish to generalise the discussion of Section
\ref{sec:dtermtransmission} to a warped space.  In particular we
consider a slice of five dimensional AdS space \cite{Randall:1999ee}
whose fifth dimension is compactified on an orbifold $S^1/Z_2$, thus
$0\le y \le \pi R$.  The orbifold has 2 fixed points $y=0$ and $y=\pi
R$ at which we place 3 branes.  The metric on the space is given by,
\be
ds^2=g_{\mu\nu}dx^\mu dx^\nu +dy^2=e^{-2\sigma(y)}\eta_{\mu\nu}dx^\mu dx^\nu +
dy^2, 
\ee
where $\mu,\nu$ run over our 4 spacetime dimensions and $\sigma(y)=k|y|$
where $k^{-1}$ is the AdS curvature length.

The equations of motion now contain factors of the metric and are,
\bea
\sqrt{-g}D^\prime+\partial_5\left(\sqrt{-g}\Sigma\right)+\sqrt{-g}\delta(y-\pi
R)\sqrt{2\pi R} D_{\pi R}
&=&0\\
\partial_5 D^\prime&=&0 \label{eq:eom}.
\eea
As before these can be solved by using the periodicity of $\Sigma$.  Due to
the factors of the metric $\Sigma$ no longer has constant slope in the bulk,
instead it has an exponential profile (see Figure~\ref{fig:sigmawarped}),
\bea
D^\prime &=& -\frac{4k\sqrt{2\pi R}}{e^{8k\pi R}-1} D_{\pi R}
\\
\Sigma(y)&=&\frac{D^\prime }{4k} (1-e^{4k |y|})\frac{y}{|y|}.
\eea

Note that just as in the flat case, the dangerous operator
(\ref{eq:dangerous}) which would transmit soft masses to scalars
dynamically vanishes by virtue of the equations of motion,
specifically (\ref{eq:eom}).

Using the inspiration of the AdS/CFT correspondance one can attempt to
construct models in which the hidden sector are the IR dynamics of a
CFT and the visible and hidden sectors are conformally sequestered.
The $U(1)$ in the bulk would then corresond to an exact global
symmetry of the conformal symmetry which is weakly gauged.

\section{Conclusions}\label{sec:conclusion}

For AMSB to be the dominant source of SUSY breaking the Kahler
potential has to take on a ``sequestered'' form.  This can be achieved
by separating the SUSY breaking dynamics from the SM, either
geometrically or through large anomalous dimensions.  We presented
explicit models where the sequestered SUSY breaking dynamics involve
a gauged $U(1)$ leading to a supersoft operator for hypercharge.  We
demonstrated how through this operator the singlet introduced in the
NMSSM to solve the $\mu$ problem may be used to solve the tachyon
problem.  These two competing contributions to SUSY breaking are the
same size not through choice but \emph{dynamics}.  In the visible
sector the field content is simply that of the NMSSM.  The spectrum is
similar to that of AMSB with an additional positive shift of all the
scalar masses proportional to the square of their hypercharge.  The
theory remains UV insensitive and the bino has a large Dirac mass.

The supersoft operator treats the real and imaginary parts of the
singlet differently and in our minimal model this leads to spontaneous
CP violation in the Higgs sector.  We found that the predicted size of
the electron EDM is right at the present day constraints.  Along with
a Dirac bino this is one of the most interesting signatures of our
model.  The significant CP violation in the Higgs sector will cause
significant deviations from conventional Higgs phenomenology and
warrants further investigation.

The minimal model we present is very tightly constrained since the
same field that is involved in EWSB is also involved in fixing the
tachyon problem of AMSB.  This results in the amount of
viable parameter space being quite small and it appears to be tuned below
the percent level.  However, there are generalisations
of the model with less minimal field content where the two effects are
separated, we intend to investigate these further elsewhere
\cite{future}.  For instance\footnote{We thank Neal Weiner for
discussions on this issue.}, by adding an additional singlet and
gauging B-L the supersoft operator could again be used to lift the
slepton squared masses while the $\mu$ problem is solved in an independent
sector.

\medskip
{\bf Acknowledgements}\\

We thank Hitoshi Murayama, Ann Nelson, Raman Sundrum and Neal
Weiner for conversations.  P.J.F. would like to thank
the JHU theory group for hospitality while this work was initiated.
The work of P.J.F. is supported in part by the U.S. Department of
Energy.  The work of D.E.K. is supported by the National Science 
Foundation and the Department of Energy's Outstanding Junior
Investigator Program.

\bibliography{amsbss_refs}

\providecommand{\href}[2]{#2}\begingroup\raggedright\begin{thebibliography}{10}

\bibitem{Randall:1998uk}
L.~Randall and R.~Sundrum, {\it Out of this world supersymmetry breaking},
  {\em Nucl. Phys.} {\bf B557} (1999) 79--118,
  [\href{http://xxx.lanl.gov/abs/hep-th/9810155}{{\tt hep-th/9810155}}].

\bibitem{Giudice:1998xp}
G.~F. Giudice, M.~A. Luty, H.~Murayama, and R.~Rattazzi, {\it Gaugino mass
  without singlets},  {\em JHEP} {\bf 12} (1998) 027,
  [\href{http://xxx.lanl.gov/abs/hep-ph/9810442}{{\tt hep-ph/9810442}}].

\bibitem{Chacko:1999am}
Z.~Chacko, M.~A. Luty, I.~Maksymyk, and E.~Ponton, {\it Realistic
  anomaly-mediated supersymmetry breaking},  {\em JHEP} {\bf 04} (2000) 001,
  [\href{http://xxx.lanl.gov/abs/hep-ph/9905390}{{\tt hep-ph/9905390}}].

\bibitem{Katz:1999uw}
E.~Katz, Y.~Shadmi, and Y.~Shirman, {\it Heavy thresholds, slepton masses and
  the mu term in anomaly mediated supersymmetry breaking},  {\em JHEP} {\bf 08}
  (1999) 015, [\href{http://xxx.lanl.gov/abs/hep-ph/9906296}{{\tt
  hep-ph/9906296}}].

\bibitem{Jack:2000cd}
I.~Jack and D.~R.~T. Jones, {\it Fayet-iliopoulos d-terms and anomaly mediated
  supersymmetry breaking},  {\em Phys. Lett.} {\bf B482} (2000) 167--173,
  [\href{http://xxx.lanl.gov/abs/hep-ph/0003081}{{\tt hep-ph/0003081}}].

\bibitem{Carena:2000ad}
M.~Carena, K.~Huitu, and T.~Kobayashi, {\it Rg-invariant sum rule in a
  generalization of anomaly mediated susy breaking models},  {\em Nucl. Phys.}
  {\bf B592} (2001) 164--182,
  [\href{http://xxx.lanl.gov/abs/hep-ph/0003187}{{\tt hep-ph/0003187}}].

\bibitem{Allanach:2000gu}
B.~C. Allanach and A.~Dedes, {\it R-parity violating anomaly mediated
  supersymmetry breaking},  {\em JHEP} {\bf 06} (2000) 017,
  [\href{http://xxx.lanl.gov/abs/hep-ph/0003222}{{\tt hep-ph/0003222}}].

\bibitem{Chacko:2000wq}
Z.~Chacko, M.~A. Luty, E.~Ponton, Y.~Shadmi, and Y.~Shirman, {\it The gut scale
  and superpartner masses from anomaly mediated supersymmetry breaking},  {\em
  Phys. Rev.} {\bf D64} (2001) 055009,
  [\href{http://xxx.lanl.gov/abs/hep-ph/0006047}{{\tt hep-ph/0006047}}].

\bibitem{Kaplan:2000jz}
D.~E. Kaplan and G.~D. Kribs, {\it Gaugino-assisted anomaly mediation},  {\em
  JHEP} {\bf 09} (2000) 048,
  [\href{http://xxx.lanl.gov/abs/hep-ph/0009195}{{\tt hep-ph/0009195}}].

\bibitem{Arkani-Hamed:2000xj}
N.~Arkani-Hamed, D.~E. Kaplan, H.~Murayama, and Y.~Nomura, {\it Viable
  ultraviolet-insensitive supersymmetry breaking},  {\em JHEP} {\bf 02} (2001)
  041, [\href{http://xxx.lanl.gov/abs/hep-ph/0012103}{{\tt hep-ph/0012103}}].

\bibitem{Jones:2001iv}
D.~R.~T. Jones, {\it Anomaly mediated supersymmetry breaking, d-terms and r-
  symmetry},  \href{http://xxx.lanl.gov/abs/hep-ph/0101159}{{\tt
  hep-ph/0101159}}.

\bibitem{Abe:2001cg}
N.~Abe, T.~Moroi, and M.~Yamaguchi, {\it Anomaly-mediated supersymmetry
  breaking with axion},  {\em JHEP} {\bf 01} (2002) 010,
  [\href{http://xxx.lanl.gov/abs/hep-ph/0111155}{{\tt hep-ph/0111155}}].

\bibitem{Chacko:2001jt}
Z.~Chacko and M.~A. Luty, {\it Realistic anomaly mediation with bulk gauge
  fields},  {\em JHEP} {\bf 05} (2002) 047,
  [\href{http://xxx.lanl.gov/abs/hep-ph/0112172}{{\tt hep-ph/0112172}}].

\bibitem{Nelson:2001ji}
A.~E. Nelson and N.~J. Weiner, {\it Gauge/anomaly syzygy and generalized brane
  world models of supersymmetry breaking},  {\em Phys. Rev. Lett.} {\bf 88}
  (2002) 231802, [\href{http://xxx.lanl.gov/abs/hep-ph/0112210}{{\tt
  hep-ph/0112210}}].

\bibitem{Okada:2002mv}
N.~Okada, {\it Positively-deflected anomaly mediation},  {\em Phys. Rev.} {\bf
  D65} (2002) 115009, [\href{http://xxx.lanl.gov/abs/hep-ph/0202219}{{\tt
  hep-ph/0202219}}].

\bibitem{Nelson:2002sa}
A.~E. Nelson and N.~T. Weiner, {\it Extended anomaly mediation and new physics
  at 10-tev},  \href{http://xxx.lanl.gov/abs/hep-ph/0210288}{{\tt
  hep-ph/0210288}}.

\bibitem{Jack:2003qg}
I.~Jack and D.~R.~T. Jones, {\it Yukawa textures and anomaly mediated
  supersymmetry breaking},  {\em Nucl. Phys.} {\bf B662} (2003) 63--88,
  [\href{http://xxx.lanl.gov/abs/hep-ph/0301163}{{\tt hep-ph/0301163}}].

\bibitem{Anoka:2003kn}
O.~C. Anoka, K.~S. Babu, and I.~Gogoladze, {\it Tev-scale horizontal symmetry
  and the slepton mass problem of anomaly mediation},  {\em Nucl. Phys.} {\bf
  B686} (2004) 135--154, [\href{http://xxx.lanl.gov/abs/hep-ph/0312176}{{\tt
  hep-ph/0312176}}].

\bibitem{Kitano:2004zd}
R.~Kitano, G.~D. Kribs, and H.~Murayama, {\it Electroweak symmetry breaking via
  uv insensitive anomaly mediation},  {\em Phys. Rev.} {\bf D70} (2004) 035001,
  [\href{http://xxx.lanl.gov/abs/hep-ph/0402215}{{\tt hep-ph/0402215}}].

\bibitem{Shafi:2004cf}
Q.~Shafi and Z.~Tavartkiladze, {\it Sparticle masses, mu problem and anomaly
  mediated supersymmetry breaking},
  \href{http://xxx.lanl.gov/abs/hep-ph/0408156}{{\tt hep-ph/0408156}}.

\bibitem{Sundrum:2004un}
R.~Sundrum, {\it 'gaugomaly' mediated susy breaking and conformal
  sequestering},  \href{http://xxx.lanl.gov/abs/hep-th/0406012}{{\tt
  hep-th/0406012}}.

\bibitem{Ibe:2004gh}
M.~Ibe, R.~Kitano, and H.~Murayama, {\it A viable supersymmetric model with uv
  insensitive anomaly mediation},
  \href{http://xxx.lanl.gov/abs/hep-ph/0412200}{{\tt hep-ph/0412200}}.

\bibitem{Fox:2002bu}
P.~J. Fox, A.~E. Nelson, and N.~Weiner, {\it Dirac gaugino masses and supersoft
  supersymmetry breaking},  {\em JHEP} {\bf 08} (2002) 035,
  [\href{http://xxx.lanl.gov/abs/hep-ph/0206096}{{\tt hep-ph/0206096}}].

\bibitem{Luty:2001jh}
M.~A. Luty and R.~Sundrum, {\it Supersymmetry breaking and composite extra
  dimensions},  {\em Phys. Rev.} {\bf D65} (2002) 066004,
  [\href{http://xxx.lanl.gov/abs/hep-th/0105137}{{\tt hep-th/0105137}}].

\bibitem{Luty:2001zv}
M.~Luty and R.~Sundrum, {\it Anomaly mediated supersymmetry breaking in four
  dimensions, naturally},  {\em Phys. Rev.} {\bf D67} (2003) 045007,
  [\href{http://xxx.lanl.gov/abs/hep-th/0111231}{{\tt hep-th/0111231}}].

\bibitem{Jack:1999fa}
I.~Jack and D.~R.~T. Jones, {\it Quasi-infra-red fixed points and
  renormalization group invariant trajectories for non-holomorphic soft
  supersymmetry breaking},  {\em Phys. Rev.} {\bf D61} (2000) 095002,
  [\href{http://xxx.lanl.gov/abs/hep-ph/9909570}{{\tt hep-ph/9909570}}].

\bibitem{Carena:2002bb}
M.~Carena, J.~R. Ellis, S.~Mrenna, A.~Pilaftsis, and C.~E.~M. Wagner, {\it
  Collider probes of the mssm higgs sector with explicit cp violation},  {\em
  Nucl. Phys.} {\bf B659} (2003) 145--178,
  [\href{http://xxx.lanl.gov/abs/hep-ph/0211467}{{\tt hep-ph/0211467}}].

\bibitem{Graesser:2001ec}
M.~Graesser and S.~Thomas, {\it Supersymmetric relations among electromagnetic
  dipole operators},  {\em Phys. Rev.} {\bf D65} (2002) 075012,
  [\href{http://xxx.lanl.gov/abs/hep-ph/0104254}{{\tt hep-ph/0104254}}].

\bibitem{Abel:2001vy}
S.~Abel, S.~Khalil, and O.~Lebedev, {\it Edm constraints in supersymmetric
  theories},  {\em Nucl. Phys.} {\bf B606} (2001) 151--182,
  [\href{http://xxx.lanl.gov/abs/hep-ph/0103320}{{\tt hep-ph/0103320}}].

\bibitem{Regan:2002ta}
B.~C. Regan, E.~D. Commins, C.~J. Schmidt, and D.~DeMille, {\it New limit on
  the electron electric dipole moment},  {\em Phys. Rev. Lett.} {\bf 88} (2002)
  071805.

\bibitem{Harris:1999jx}
P.~G. Harris {\em et.~al.}, {\it New experimental limit on the electric dipole
  moment of the neutron},  {\em Phys. Rev. Lett.} {\bf 82} (1999) 904--907.

\bibitem{Weinberg:1989dx}
S.~Weinberg, {\it Larger higgs exchange terms in the neutron electric dipole
  moment},  {\em Phys. Rev. Lett.} {\bf 63} (1989) 2333.

\bibitem{Chacko:2004mi}
Z.~Chacko, P.~J. Fox, and H.~Murayama, {\it Localized supersoft supersymmetry
  breaking},  {\em Nucl. Phys.} {\bf B706} (2005) 53--70,
  [\href{http://xxx.lanl.gov/abs/hep-ph/0406142}{{\tt hep-ph/0406142}}].

\bibitem{Pomarol:1999ie}
A.~Pomarol and R.~Rattazzi, {\it Sparticle masses from the superconformal
  anomaly},  {\em JHEP} {\bf 05} (1999) 013,
  [\href{http://xxx.lanl.gov/abs/hep-ph/9903448}{{\tt hep-ph/9903448}}].

\bibitem{Dine:1996ag}
M.~Dine, A.~E. Nelson, Y.~Nir, and Y.~Shirman, {\it New tools for low-energy
  dynamical supersymmetry breaking},  {\em Phys. Rev.} {\bf D53} (1996)
  2658--2669, [\href{http://xxx.lanl.gov/abs/hep-ph/9507378}{{\tt
  hep-ph/9507378}}].

\bibitem{Nelson:2000sn}
A.~E. Nelson and M.~J. Strassler, {\it Suppressing flavor anarchy},  {\em JHEP}
  {\bf 09} (2000) 030, [\href{http://xxx.lanl.gov/abs/hep-ph/0006251}{{\tt
  hep-ph/0006251}}].

\bibitem{Arkani-Hamed:2001tb}
N.~Arkani-Hamed, T.~Gregoire, and J.~Wacker, {\it Higher dimensional
  supersymmetry in 4d superspace},  {\em JHEP} {\bf 03} (2002) 055,
  [\href{http://xxx.lanl.gov/abs/hep-th/0101233}{{\tt hep-th/0101233}}].

\bibitem{Randall:1999ee}
L.~Randall and R.~Sundrum, {\it A large mass hierarchy from a small extra
  dimension},  {\em Phys. Rev. Lett.} {\bf 83} (1999) 3370--3373,
  [\href{http://xxx.lanl.gov/abs/hep-ph/9905221}{{\tt hep-ph/9905221}}].

\bibitem{future}
L.~Carpenter, P.~J. Fox, D.~E. Kaplan, and N.~Weiner {\em In preparation}.

\end{thebibliography}\endgroup
\bibliographystyle{JHEP}
  
\renewcommand{\theequation}{A.\arabic{equation}}
  \setcounter{equation}{0}  
  \section*{APPENDIX}  

Here we list the anomalous dimensions and beta functions relevant for
calculating anomaly mediated soft masses in the NMSSM.

\bea\label{eq:anomdims}
\gamma_{H_u} &=& \frac{1}{16\pi^2}\left(3g_2^2+g_Y^2-2\lambda_S^2-6y_t^2\right)\\
\gamma_{H_d} &=& \frac{1}{16\pi^2}\left(3g_2^2+g_Y^2-2\lambda_S^2\right)\\
\gamma_Q &=&\frac{1}{16\pi^2}\left(\frac{16}{3}g_3^2+3g_2^2+\frac{1}{9}g_Y^2
  -2y_t^2\right) \\
\gamma_{\bar{U}} &=&\frac{1}{16\pi^2}\left(\frac{16}{3}g_3^2
  +\frac{16}{9}g_Y^2 -4y_t^2\right)\\
\gamma_{\bar{D}} &=& \frac{1}{16\pi^2}\left(\frac{16}{3}g_3^2 + 
\frac{4}{9}g_Y^2\right)\\
\gamma_L &=& \frac{1}{16\pi^2}\left(3g_2^2+g_Y^2\right)\\
\gamma_{\bar{E}} &=& \frac{1}{4\pi^2}g_Y^2\\
\gamma_S&=&-\frac{1}{4\pi^2}(\lambda_S^2+\kappa^2) 
\eea 

\be\
\frac{d g_3}{dt}=-\frac{3}{16\pi^2}g_3^3, \ \ \ 
\frac{dg_2}{dt}=\frac{1}{16\pi^2}g_2^3, \ \ \ 
\frac{d g_Y}{dt}=\frac{11}{16\pi^2}g_Y^3
\ee
\bea\label{eq:potentialbetas}
\frac{d\lambda}{dt}&=&\frac{\lambda_S}{16\pi^2} 
\left(4\lambda_S^2+2\kappa^2+3y_t^2-3g_2^2-g_Y^2\right)\\
\frac{dy_t}{dt}&=&\frac{y_t}{16\pi^2}\left(
  6y_t^2+\lambda_S^2-\frac{16}{3}g_3^2 -3g_2^2-\frac{13}{9}g_Y^2\right)\\
\frac{d\kappa}{dt}&=&\frac{3\kappa}{8\pi^2}\left(\lambda^2+\kappa^2\right)
\eea
These set of equations completely determine the one-loop anomaly
mediated contributions to the soft parameters.

\end{document}